\documentclass[conference]{IEEEtran}
\IEEEoverridecommandlockouts

\usepackage{amsmath,amssymb,amsfonts}
\usepackage{booktabs}
\usepackage{graphicx}
\usepackage{multirow}
\usepackage{url}
\usepackage{xcolor}
\usepackage[hidelinks]{hyperref}
\usepackage[nointegrals]{wasysym}
\usepackage[absolute,overlay]{textpos}
\usepackage[authormarkup=none]{changes}
\definechangesauthor[name={Revision}, color=blue]{rc}

\newcommand{\cmark}{{\normalfont\small\CIRCLE}}
\newcommand{\pmark}{{\normalfont\small\LEFTcircle}}
\newcommand{\xmark}{{\normalfont\small\Circle}}
\newcommand{\lib}[1]{\texttt{#1}}
\newcommand{\Fone}{\ensuremath{F_1}}

\begin{document}

\title{\textsc{Real Package, Wrong Controller: Static Auditing of Vendor--Protocol
Mismatches in LLM-Generated PLC Code}}

\author{
    \IEEEauthorblockN{
        S M Zia Ur Rashid\textsuperscript{1,}, 
        Sazidul Islam Hira\textsuperscript{2}, and 
        Suman Rath\textsuperscript{1}
    }
    \IEEEauthorblockA{
        \textsuperscript{1}Department of Electrical and Computer Engineering, The University of Tulsa, Tulsa, OK, USA\\
        \textsuperscript{2}Department of Computer Science and Engineering, International Islamic University Chittagong, Chittagong, Bangladesh\\
        Email: \textsuperscript{1}\{ziaur-rashid, suman-rath\}@utulsa.edu, \textsuperscript{2}sazidulislam.mail@gmail.com
    }
}

% \IEEEpubid{\makebox[\columnwidth]{979-8-3195-0171-4/26/\$31.00~\copyright2026 IEEE\hfill}\hspace{\columnsep}\makebox[\columnwidth]{ }}
\maketitle
\IEEEpubidadjcol
\begin{textblock*}{18cm}(1.8cm,0.62cm)
% \noindent\footnotesize 2026 29th International Conference on Computer and
% Information Technology (ICCIT)\\
% 18--20 December 2026, Cox's Bazar, Bangladesh.
\end{textblock*}

\begin{abstract}
Language models can generate Python that drives industrial protocols on
programmable logic controllers, yet existing code-grounding checks ask whether
an imported package or symbol exists rather than whether it suits the declared
industrial target. We define \emph{metadata-conditioned, import-level
vendor--protocol incompatibility} and build \textsc{Proteus-OT}, a label-free
static verifier over a version-pinned graph of documented bindings. Across 883
generated scripts, 180 carry an import-level mismatch against the declared
target (20.4\%); restricting to
single-vendor targets gives 16.3\%, and additionally requiring the offending
library to reach a protocol communication call gives
14.2\%, which we read as conditional on the corpus
metadata rather than on verified intent. None of the 28 adjudicable human
control files carries one, a sample too small to estimate a prevalence. Within
the balanced subset, 96.6\% of mismatching scripts
import packages and call methods that exist, so an existence baseline does not
represent them, and every positive ships with an inspectable witness. We also
audit machine-versus-human classification on the same corpus, where character
$n$-grams reach macro-F1 1.000 under task, model and vendor-disjoint
partitions and survive comment removal, an AST round trip and identifier
anonymisation. The trailing newline present in 95.2\% of human files and in none
of the generated ones separates the classes on its own. Vendor--protocol
consistency is a useful static audit target, whereas provenance classification
on repository-labelled corpora is not authorship detection.
\end{abstract}


\section{Introduction}
\label{sec:intro}

Programmable logic controllers sit where computation meets physical process,
which is why they have been decisive in the most consequential intrusions on
record~\cite{langner2011,johnson2017}. Mounting such an operation has
historically demanded process knowledge, vendor-specific protocol familiarity
and time with real hardware. Cook et al.~\cite{cook2026} report that
general-purpose language models supply part of that combination, generating PLC
attack scripts from prompts and validating them against live controllers, so
model assistance may lower the barrier to entry.

We ask the defensive question of whether such code can be checked in a form an
engineer can act on. Existing tooling reasons about token style or package
existence~\cite{spracklen2025}, and neither engages with what makes an
operational-technology (OT) script wrong. Industrial libraries are
vendor-bound, so \lib{snap7} speaks Siemens' S7comm and \lib{pycomm3} speaks
Rockwell's EtherNet/IP--CIP, yet a model prompted for a Schneider Modicon device
will often import one of them regardless, producing code that parses and calls
real methods while the documented bindings do not list that package for the
declared target. Existence checkers confirm the package and the symbol, then
stop. The gap we address is the
absence, across the venues we surveyed, of any static, pre-execution analysis
that reasons about the vendor--protocol suitability of the attack script itself
rather than about style, package existence, or run-time network behaviour.

We formalise this problem as a metadata-conditioned, import-level
incompatibility and build \textsc{Proteus-OT} around it, a static and label-free
auditor for operational-technology scripts that pairs a typed graph of public
bindings carrying per-edge sources with a verifier emitting a twelve-component
readable vector, where every positive carries an inspectable witness. We also
examined whether the same grounded features help a machine-versus-human
classifier separate the two provenance classes. That earlier experiment reported
a 4.2 macro-\Fone{} point improvement, which we now retract, because the
component meant to flag hallucinated packages was firing on intra-repository
sibling imports and so keyed on repository structure rather than authorship. More
fundamentally, provenance classification proves degenerate on corpora of this
construction, and we report the corrected, non-significant gap together with the
artifact analysis that explains it, since an improvement over a defective
baseline is not a contribution.

This paper makes four contributions.
\begin{itemize}
\item We define a metadata-conditioned, import-level vendor--protocol mismatch
predicate over a version-pinned binding graph, decidable from source with no
labels or trained model, where every positive carries an inspectable witness.
\item We audit 883 machine-generated scripts, reporting a 14.2\% use-level rate
against a 20.4\% inclusive upper bound, with 96.6\% of flagged scripts falling
outside the evaluated existence baseline.
\item We show that machine-versus-human provenance classification on
repository-labelled corpora separates the classes by
construction rather than authorship, and reduce the confound to a single serialisation byte.
\item We bound recall by graph coverage and evasion by measurement, and release
the binding graph, evidence table, adjudication worksheet, splits, environment
and pipeline.
\end{itemize}

The remainder of the paper proceeds as follows.
Section~\ref{sec:related} positions the work against generation,
package-existence and evaluation-hygiene literature.
Section~\ref{sec:method} presents the threat model, the mismatch predicate and
the verifier. Section~\ref{sec:results} reports the audit and the provenance
negative result. Section~\ref{sec:discussion} discusses which claims survive,
and Section~\ref{sec:conclusion} concludes.

\section{Related Work}
\label{sec:related}

Models produce phishing material, scanners and simple exploits under mild
prompting~\cite{benmoshe2022}, and later systems automate larger stretches of
the intrusion lifecycle~\cite{xu2024,iturbe2024}. Cook et al.~\cite{cook2026}
extend this to OT, evaluating five models on PLC attack techniques from ATT\&CK
for ICS~\cite{mitre_ics} across three vendors and three levels of supplied
target information, and their corpus is our positive class. All of this work
studies generation rather than defence.

Code models hallucinate~\cite{liu2024,tian2024}, most relevantly by importing
packages that do not exist, at rates sufficient to support supply-chain attacks
on the invented names~\cite{spracklen2025}. The unit of analysis there is
\emph{existence}, which is the wrong question in OT, where the problematic
scripts import entirely real packages belonging to another vendor. Our
contribution swaps the grounding oracle from ``does this symbol exist'' to ``do
the documented bindings list this library as supporting the declared target''.
Defensive ICS research meanwhile concentrates on network- and process-level
detection at run time, with adjacent work on model reliability for legitimate
PLC logic~\cite{fakih2024}, which targets correctness of generated control logic
rather than pre-execution auditing of an attack script. In the venues we
surveyed we found no static, pre-execution analysis of the attack script itself.

Our second result connects to work on evaluation hygiene. It builds on recurring
pitfalls in security machine learning~\cite{arp2022}, malware results that do
not survive realistic partitions~\cite{pendlebury2019}, duplication inflating
measured accuracy on code beyond what exact-hash deduplication
removes~\cite{allamanis2019}, and leakage as a cause of
irreproducibility~\cite{kapoor2023}. The closest analogue is linguistic, where
annotation artifacts let hypothesis-only models beat chance on inference
benchmarks~\cite{gururangan2018}, as a newline-only model does on ours. For
generated text, reliable detection faces fundamental limits~\cite{sadasivan2023}
and deployed detectors encode confounds~\cite{liang2023}.
Table~\ref{tab:related} positions the work along OT domain, whether the method
detects rather than only generates, vendor semantics, label-free operation,
leakage control and an artifact audit, adding the last two columns we find
missing from generated-code detection work, and its marks report what each paper
states rather than what its method could in principle support.

\begin{table}[t]
\centering
\caption{Positioning against representative prior work, where \cmark\ marks a
property present, \pmark\ partial and \xmark\ absent.}
\label{tab:related}
\footnotesize
\setlength{\tabcolsep}{2.5pt}
\begin{tabular}{lcccccc}
\toprule
\textbf{Work} &
\rotatebox{55}{OT domain} &
\rotatebox{55}{Detects} &
\rotatebox{55}{Vendor sem.} &
\rotatebox{55}{Label-free} &
\rotatebox{55}{Leakage ctrl.} &
\rotatebox{55}{Artifact audit} \\
\midrule
Ben-Moshe et al.~\cite{benmoshe2022}    & \xmark & \xmark & \xmark & \xmark & \xmark & \xmark \\
AutoAttacker~\cite{xu2024}              & \xmark & \xmark & \xmark & \xmark & \xmark & \xmark \\
Cook et al.~\cite{cook2026}             & \cmark & \xmark & \pmark & \xmark & \xmark & \xmark \\
Spracklen et al.~\cite{spracklen2025}   & \xmark & \cmark & \xmark & \pmark & \pmark & \xmark \\
Fakih et al.~\cite{fakih2024}           & \cmark & \xmark & \xmark & \xmark & \xmark & \xmark \\
Pendlebury et al.~\cite{pendlebury2019} & \xmark & \cmark & \xmark & \xmark & \cmark & \pmark \\
Allamanis~\cite{allamanis2019}          & \xmark & \xmark & \xmark & \xmark & \cmark & \cmark \\
\textbf{Ours}                           & \cmark & \cmark & \cmark & \cmark & \cmark & \cmark \\
\bottomrule
\end{tabular}
\end{table}

\section{Methodology}
\label{sec:method}

\subsection{Threat Model and Objectives}

We adopt the OT threat model of~\cite{cook2026}, in which an adversary prompts a
model and receives a Python script intended to speak an industrial protocol to a
controller. Following the SANS ICS kill chain~\cite{assante2015} we address the
payload, not the foothold. The defender inspects a script statically, before any
packet reaches equipment.

Let $s$ be a script, $\mathcal{V}$ the vendor set, and $v(s)\in\mathcal{V}$ the
declared target where one exists ($\bot$ otherwise). We separate two objectives.
RQ1 asks, given $s$ and $v(s)\neq\bot$,
whether $s$ imports a library that the documented bindings do not list as
supporting $v(s)$, and emits a witness a human can inspect without labels. RQ2
asks, given labels $y_i$ that indicate machine or human authorship, to learn
$f\!:\!s\mapsto\{0,1\}$, where the labels must mean what they claim. The
deterministic RQ1 result is not evidence for RQ2. RQ1 is our contribution, and
RQ2 is reported as a case study in why such corpora do not support the claim.

When the predicate fires, it establishes only that, under the encoded bindings
and the declared metadata, the imported library is not documented as supporting
the target. Firing does \emph{not} establish that the script cannot communicate
with the device, that no gateway or translation layer mediates, that the library
lacks an undocumented mode, that the declared target is correct, or that the
model hallucinated rather than erring otherwise.

\subsection{Binding Graph and Mismatch Predicate}

The backbone is a typed graph over
$\mathcal{N}=\mathcal{V}\cup\mathcal{P}\cup\mathcal{L}\cup\mathcal{S}$
(vendors, protocols, libraries, symbols). Modelling the library--protocol
binding as a function $\lambda:\mathcal{L}\to\mathcal{P}$ is too strong, since a
package may implement several protocols or expose vendor-specific extensions
beside generic ones. We use a relation and make
vendor validity conditional on a context $\theta$ carrying package version,
transport and device family,
% \begin{equation}
% \Lambda\subseteq\mathcal{L}\times\mathcal{P},\;
% \mathrm{Protocols}(\ell)=\{p:(\ell,p)\in\Lambda\},\;
% \nu:\mathcal{P}\times\Theta\to 2^{\mathcal{V}\cup\{\ast\}}
% \label{eq:lambda}
% \end{equation}
\begin{equation}
\begin{gathered}
\Lambda \subseteq \mathcal{L} \times \mathcal{P}, \quad
\mathrm{Protocols}(\ell) = \{p : (\ell, p) \in \Lambda\}, \\
\nu : \mathcal{P} \times \Theta \to 2^{\mathcal{V} \cup \{\ast\}}
\end{gathered}
\label{eq:lambda}
\end{equation}
where $\ast$ marks a vendor-neutral protocol. Validity is \emph{existential}
over the protocols a library implements,
\begin{equation}
\begin{gathered}
\mathrm{Valid}(\ell,v,\theta) \equiv \exists\,p\in\mathrm{Protocols}(\ell) : \\
\bigl[\ast\in\nu(p,\theta)\vee v\in\nu(p,\theta)\bigr]
\end{gathered}
\label{eq:valid}
\end{equation}
Libraries offering any protocol that is vendor-neutral, or valid for the
declared vendor, are therefore never witnesses. Validity here turns on vendor
identity, a necessary but not sufficient compatibility check that
Section~\ref{sec:scope} bounds. The witness set and the three-valued predicate
follow, with $L(s)$ the OT libraries a script imports,
\begin{equation}
M(s,v)=\{\ell\in L(s):\neg\,\mathrm{Valid}(\ell,v,\theta)\}
\label{eq:witness}
\end{equation}
\begin{equation}
g_{\text{xv}}(s)=
\begin{cases}
1, & v(s)\neq\bot \wedge M(s,v(s))\neq\emptyset\\
0, & v(s)\neq\bot \wedge M(s,v(s))=\emptyset\\
\varnothing, & v(s)=\bot
\end{cases}
\label{eq:gxv}
\end{equation}
The undefined case fixes the denominator, so rates are taken over
$\{s:v(s)\neq\bot\}$, 883 of 883 machine scripts but only 28 of 190 human ones.

The predicate is cheap to decide. With a finite AST, recognised import set and
binding graph and fixed target and context metadata, $g_{\text{xv}}$ is decidable
in time linear in AST size plus imported OT libraries, a bounded lookup per
import, and concerns a syntactic metadata-conditioned predicate rather than
protocol compatibility, runtime reachability or exploitability. Here $\theta$ is
fixed (pinned versions, default transport, no device family), so the
context-conditioning of Eqs.~\eqref{eq:lambda} and~\eqref{eq:valid} is specified
for future use and does not bear on the results.

\subsection{Verifier and Features}

The verifier parses $s$ to an AST, never executing it, excludes relative imports
as package-internal siblings, and emits $g(s)\in\mathbb{R}^{12}$
(Table~\ref{tab:features}), partitioned as $g=g_{\text{gen}}\oplus g_{\text{ot}}$,
where the existence-only part $g_{\text{gen}}$ (column G) holds only what an
existence detector~\cite{spracklen2025} could compute and the OT part isolates
what grounding contributes.

One component required correction. Any import absent from a whitelist previously
counted as invented, which on a flattened corpus mount fired on intra-repository
modules and keyed on repository structure, so we now build, per repository $r$,
the set $M_r$ of module names genuinely present there,
\begin{equation}
g_{\text{imag}}(s)=\mathbb{1}\bigl[\exists\, i\in I(s):
i\notin\mathcal{K}\wedge i\notin M_{r(s)}\bigr],
\label{eq:imag}
\end{equation}
with $\mathcal{K}$ the pinned known-package registry. Repository-local scoping
preserves genuine signal, and \lib{pyopcuapy}, occurring nowhere in the
generating corpus, still fires. Class means moved from 0.000/0.155
(machine/human) to 0.000/0.042.

\begin{table}[t]
\centering
\caption{The twelve groundedness components with their definitions and ranges.}
\label{tab:features}
\footnotesize
\setlength{\tabcolsep}{3pt}
\begin{tabular}{llp{4.15cm}}
\toprule
\textbf{Component} & \textbf{Set} & \textbf{Definition (range)} \\
\midrule
\texttt{imaginary\_library}   & G  & Eq.~\eqref{eq:imag}, import absent from registry and from $M_r$ \{0,1\} \\
\texttt{nonexistent\_symbol}  & G  & invented methods on real OT libraries, count $\ge 0$ \\
\texttt{placeholder\_comment} & G  & comment matching insert/TODO/your-code \{0,1\} \\
\texttt{redundant\_import}    & G  & duplicate top-level import \{0,1\} \\
\texttt{comment\_density}     & G  & comment lines / total lines $[0,1]$ \\
\texttt{cross\_vendor\_mismatch} & OT & Eq.~\eqref{eq:gxv} \{0,1,$\varnothing$\} \\
\texttt{port\_protocol\_conflict} & OT & declared port is a known OT port other than the protocol's \{0,1\} \\
\texttt{addressing\_violation} & OT & Modbus-style 4xxxx address with S7comm \{0,1\} \\
\texttt{imaginary\_protocol}  & OT & protocol token with no graph node \{0,1\} \\
\texttt{tag\_fabrication\_risk} & OT & tag-like literals disjoint from supplied tags, e.g.\ \texttt{"Motor7"} \{0,1\} \\
\texttt{ungrounded\_ratio}    & OT & attribute accesses absent from pinned symbol tables, e.g.\ a misspelled method $[0,1]$ \\
\texttt{n\_ot\_libs}          & -- & $|L(s)|$ \\
\bottomrule
\end{tabular}
\end{table}

\subsection{Witness Depth}

Imports are not functional dependencies, so each positive is classified by how
deeply the script uses the offending library. With alias tracking
(\texttt{import lib as x}, \texttt{from lib import X}) and one level of
assignment binding, we assign L0 when the alias is never referenced, L1 when it
is referenced but nothing is called, L2 when a library attribute is called, and
L3 when a call reaches a symbol on the protocol communication allowlist
(\texttt{connect}, \texttt{db\_read}, \texttt{read\_holding\_registers} and so
on), directly or through an object constructed from the library, which is the
strongest static evidence short of execution that the library sits on the
communication path.

\subsection{Corpus and Partitions}

The machine class is the attack-synthesis corpus of~\cite{cook2026} and the
human class is public OT repositories, comprising security tooling and the source
of the mainstream Python OT libraries, from which we keep only files importing a
real OT library. Labels are \emph{provenance} labels, not adjudicated authorship.
We recover the generating model, ATT\&CK technique and supplied-context token
from file naming to support grouped partitions.

Corpus construction is deterministic and fingerprinted. Discovered paths are
sorted before use, because enumeration order varies across runs on identical
data, and during development this moved random-CV macro-\Fone{} by 0.033 and
flipped the sign of an ablation. Every result below carries fingerprint
\texttt{1fc2fa6b677c} over 1073 files. After exact deduplication (55 files) and
normalised-token near-duplicate clustering at Jaccard $\ge 0.90$ on 5-shingles
(12 files, 1.2\%), 1006 remain, and balancing yields 336 for classification. The
audit uses the full unbalanced corpus and needs no labels.

We report five partitions, keeping the random one only to make the drop to
controlled schemes visible. These are random 5-fold, \emph{task-disjoint} by
ATT\&CK technique, \emph{leave-one-model-out} by generator,
\emph{leave-one-vendor-out} by target vendor with vendor groups over machine rows
only, and \emph{human-repository-disjoint}. Where a group maps one-to-one onto a
class we hold out that group and split the other class evenly, discarding and
counting single-class folds. Leave-one-vendor-out carries a caveat, since vendor
strings appear in identifiers and paths, so it bounds vendor-idiom transfer
rather than authorship transfer.

\subsection{Baselines and Metrics}

Against the grounded feature set we evaluate the existence-only restriction, the
ablation removing $g_{\text{ot}}$, formatting statistics only, a normalised AST
node histogram, character $n$-gram TF--IDF ($n\in[3,5]$), frozen
CodeBERT~\cite{feng2020}, and a rule-only comparator on $g_{\text{xv}}$. Every
vectoriser is fitted inside the fold, and we use
\texttt{scikit-learn}~\cite{pedregosa2011}. We report macro-\Fone{} with
per-class precision and recall, PR-AUC, ROC-AUC, Matthews correlation and
balanced accuracy, plus per-fold values with bootstrap intervals. Paired
differences use an exact sign-flip permutation test over folds with Holm
correction~\cite{holm1979} across the partition family, and
Wilcoxon~\cite{wilcoxon1945} is reported for comparability. Proportions use the
Wilson interval~\cite{wilson1927},
\begin{equation}
\frac{\hat{p}+\frac{z^2}{2n}}{1+z^2/n}
\pm\frac{z}{1+z^2/n}\sqrt{\frac{\hat{p}(1-\hat{p})}{n}+\frac{z^2}{4n^2}}
\label{eq:wilson}
\end{equation}
because several denominators are small and one proportion is exactly zero. Seeds
measure sensitivity to resampling rather than external robustness.

For the separability diagnostic let $\Phi_1$ delete comments and docstrings,
$\Phi_2$ additionally collapse whitespace, $\Phi_3$ additionally re-emit source
through an AST round trip destroying all formatting, and $\Phi_4$ additionally
anonymise local identifiers; with $\Phi_0=\mathrm{id}$ we report
$\mathcal{F}(k)=F_1^{\text{macro}}(\text{TF--IDF}\circ\Phi_k(\mathcal{D}))$.
Rungs 3 and 4 strip docstrings on the AST of the original source rather than on
the text-level stripper's output, because deleting a docstring as text can leave
an empty block that no longer parses, and that defect previously produced 42\%
self-inflicted parse failures on the human class, which were then encoded as
empty documents. We also delete the $m$ most heavily weighted features and
refit, and report the discrimination each surface statistic $\phi$ achieves
alone as $\mathrm{AUC}(\phi)$, where values near 0 and near 1 are equally
separating.

\section{Results}
\label{sec:results}

\subsection{Audit Rates and Witness Depth}
\label{sec:res-audit}

\begin{figure}[t]
\centering
\includegraphics[width=0.99\columnwidth]{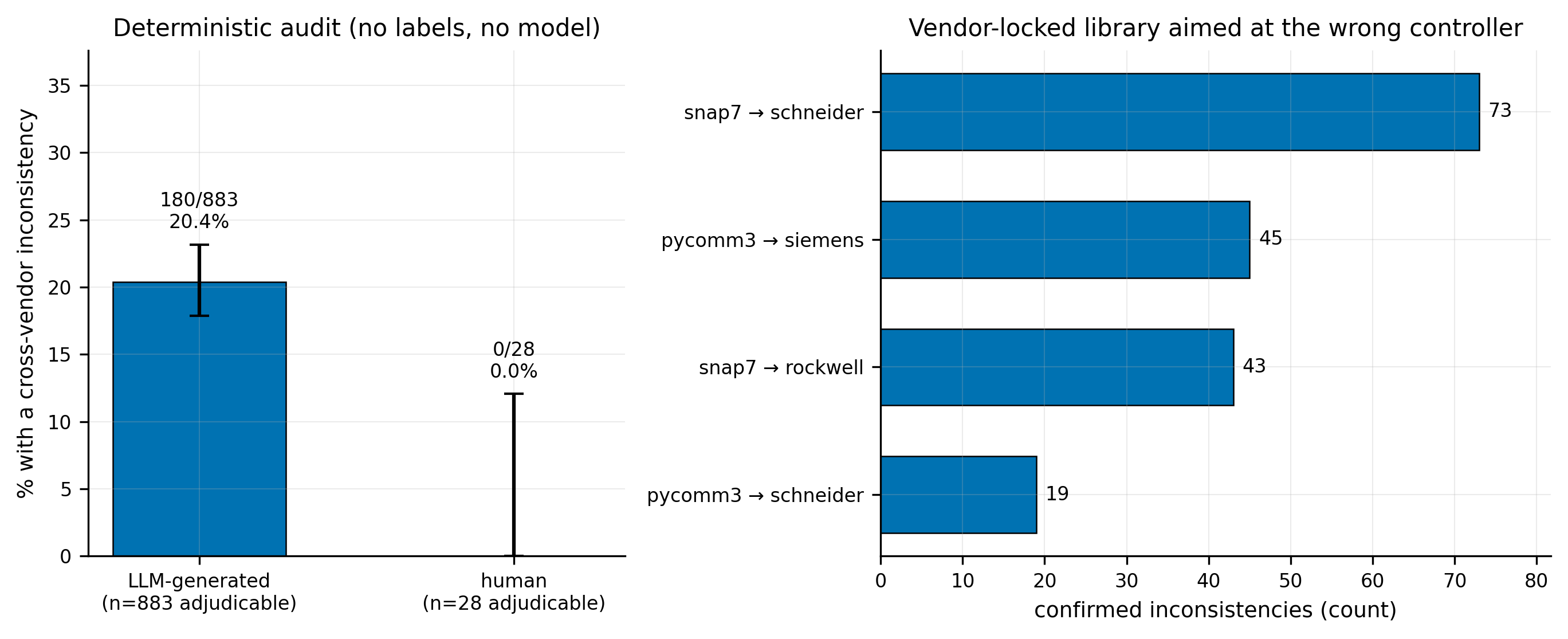}
\caption{Inclusive audit rates with Wilson intervals and library--target
pairings.}
\label{fig:audit}
\end{figure}

Table~\ref{tab:audit} and Fig.~\ref{fig:audit} give the audit over the full
unbalanced corpus. The inclusive rate is 180/883 = 20.4\% [17.9, 23.2], every
positive deriving from path-supplied metadata; 44 flagged scripts name more than
one vendor, so their target is ambiguous and excluding them gives 16.3\% [14.0,
19.0]; and requiring in addition that the offending library reach a protocol
communication call gives 14.2\% [12.0, 16.7],
our primary metadata-conditioned estimate, with the inclusive rate as an upper bound. Fig.~\ref{fig:audit} shows
the mismatch concentrated in a few library--target pairings, led by \lib{snap7}
aimed at Schneider (73 scripts) and \lib{pycomm3} aimed at Siemens (45), with
\lib{snap7} at Rockwell (43) and \lib{pycomm3} at Schneider (19) following.

Witness depth concentrates at full use, with 3 of the 180 witnesses (1.7\%) dead
imports, none referenced-but-uncalled, 19 (10.6\%) calling the library without
reaching a communication symbol, and 158 (87.8\% [82.2, 91.8]) reaching one. At
98\% live dependencies, this addresses the objection that an import witness need
not be functional. Among human controls, 28 of 190 files carry a recoverable
target and none is flagged (Wilson upper bound 12.1\%), too few to estimate a
prevalence and not evidence that the mismatch is specific to generated code. The
human class is itself skewed, since 142 of its 190 files (75\%) are library source
and tests rather than standalone attack code, so the 28 adjudicable files
over-represent code that imports its own vendor's library by construction, and the
zero count should be read as the absence of a matched control rather than as
demonstrated human care.

\begin{table}[t]
\centering
\caption{Audit over the full unbalanced corpus (fingerprint
\texttt{1fc2fa6b677c}).}
\label{tab:audit}
\footnotesize
\begin{tabular}{lrrl}
\toprule
\textbf{Estimate} & \textbf{Num.} & \textbf{Den.} & \textbf{Rate [95\% CI]} \\
\midrule
Inclusive (all path-derived)      & 180 & 883 & 20.4\% [17.9, 23.2] \\
Single-vendor target only         & 136 & 833 & 16.3\% [14.0, 19.0] \\
\textbf{Use-level confirmed (L3)} & \textbf{118} & \textbf{833} & \textbf{14.2\% [12.0, 16.7]} \\
Human control                     &   0 &  28 & \ \,0.0\% [0.0, 12.1] \\
\bottomrule
\end{tabular}
\end{table}

Because the declared target comes from file naming, we recovered it independently
from the code text and compared the two channels on the 318 scripts both can
adjudicate. The vendors agree on 84.9\% of
those scripts, but the flag disagreement is lopsided, with filename flags at
23.9\% [19.5, 28.9] against code-text at 11.6\% [8.6, 15.6], and 41 flagged only
by filename and 2 only by code text. When a script names a vendor in its own text
it usually names the vendor of the library it imported, so 14.2\% is sensitive to
which channel supplies the target and should be read as conditional on the corpus
metadata rather than on independently verified intent.

\subsection{The Existence Baseline}

The existence baseline flags a script if an imported top-level package is absent
from the pinned registry, or a called attribute is absent from the introspected
symbol table of an imported OT library. Within the balanced corpus, 29 of 168
machine scripts carry an OT-semantic flag, and the baseline is silent on 28 and
fires on 1, plus 4 scripts with no OT-semantic flag, so 96.6\% [82.8, 99.4] are
not represented by it. The claim concerns the evaluated baseline rather than
every conceivable existence-based tool, since adding vendor--protocol edges would
make a tool no longer existence-based.

\subsection{Provenance Classification as a Negative Result}
\label{sec:res-prov}

Table~\ref{tab:main} reports every model under every partition, and the pattern
it shows is separation by construction rather than by authorship. Character
$n$-grams attain macro-\Fone{} 1.000 under random, task-disjoint,
leave-one-model-out and leave-one-vendor-out partitions and 0.994
human-repository-disjoint, above frozen CodeBERT and far above every semantic
feature set. That a pretrained code model is beaten by character $n$-grams is
hard to reconcile with the label meaning what it claims.

\begin{table}[t]
\centering
\caption{Macro-\Fone{} fold means for machine-versus-human classification, with
character $n$-gram values in bold.}
\label{tab:main}
\footnotesize
\setlength{\tabcolsep}{3pt}
\begin{tabular}{lccccc}
\toprule
\textbf{Feature space} & \textbf{Rand.} & \textbf{Task} & \textbf{Model} &
\textbf{Vendor} & \textbf{Repo} \\
\midrule
Char $n$-gram TF--IDF     & \textbf{1.000} & \textbf{1.000} & \textbf{1.000} & \textbf{1.000} & \textbf{0.994} \\
CodeBERT (frozen)~\cite{feng2020} & 0.988 & 0.988 & 0.977 & 0.978 & 0.973 \\
Formatting statistics     & 0.943 & 0.940 & 0.900 & 0.934 & 0.914 \\
AST node histogram        & 0.937 & 0.931 & 0.911 & 0.923 & 0.907 \\
\midrule
\textsc{Proteus-OT} (MLP) & 0.893 & 0.899 & 0.861 & 0.898 & 0.849 \\
\textsc{Proteus-OT} (LR)  & 0.881 & 0.880 & 0.834 & 0.892 & 0.878 \\
Ablation, no OT layer     & 0.866 & 0.853 & 0.845 & 0.856 & 0.834 \\
Existence-only            & 0.854 & 0.840 & 0.829 & 0.848 & 0.816 \\
Mismatch rule alone       & 0.500 & 0.496 & 0.565 & 0.538 & 0.426 \\
\midrule
Grounded $-$ existence    & $+0.027$ & $+0.040$ & $+0.004$ & $+0.044$ & $+0.062$ \\
Permutation $p$           & 0.250 & 0.125 & 0.813 & 0.250 & 0.125 \\
Folds                     & 5 & 5 & 5 & 3 & 4 \\
\bottomrule
\end{tabular}
\end{table}

The grounded-minus-existence contrast is consistently positive, from $+0.027$ to
$+0.062$ across partitions, and over 25 random-CV seeds it is $+0.0371$
(sd 0.0070), positive in every seed, with the OT-layer ablation at $+0.0228$
(sd 0.0070), also positive in every seed. No partition reaches significance under
an exact sign-flip permutation test ($p$ between 0.125 and 0.813; Holm-adjusted
0.375 to 1.00), which with three to five folds has very little resolution, since
$k$ folds admit only $2^{k}$ sign assignments and the three-fold partitions
cannot fall below $p=0.25$ regardless of effect size. The correct statement is
that the grounded features are consistently better than existence-only by a
small margin that this design cannot resolve statistically, and that the
comparison is made on a task the next subsection shows to be unsound. We withdraw
the earlier $+0.042$ claim and make no aggregate-accuracy claim in its place.

\subsection{Diagnosing Separability}
\label{sec:res-diag}

The tokens that separate the classes describe genre rather than authorship, with
human-class \texttt{self} (0.0\% machine / 63.7\% human), \texttt{assert}
(0.0/52.4), \texttt{pytest} (0.0/49.4), \texttt{class} (2.4/66.1) and
\texttt{async def} (0.0/31.0) set against machine-class \texttt{plc\_ip}
(95.2/0.6), \texttt{find\_plc\_ip} (57.1/0.0), \texttt{print} (97.6/24.4) and
\texttt{ip address} (85.7/5.4). These are the signatures of a test suite against
a standalone script, which is what the corpus contains.

Table~\ref{tab:diag} reports the normalisation ladder with feature deletion on
the left and the discrimination of single surface statistics on the right, and
it shows the separation is neither cosmetic nor carried by a few tokens. Every
normalisation rung retains macro-\Fone{} 1.000, including the AST round trip that
destroys all formatting and the anonymisation of every local identifier, and all
336 scripts round-trip successfully at every rung, so no result is contaminated
by transform failures. Deleting the 1000 most heavily weighted features leaves
0.9913, yet a single statistic still suffices, since whether a file ends with a
newline attains AUC 0.024, with no machine-generated script ending in a newline
and 95.2\% of human files doing so, a serialisation artifact of the generating
corpus exploitable with no model at all.

\begin{table}[t]
\centering
\caption{Diagnostic on the task-disjoint partition.}
\label{tab:diag}
\footnotesize
\setlength{\tabcolsep}{3pt}
\begin{tabular}{lc@{\hskip 1em}lc}
\toprule
\textbf{Normalisation} & $\mathcal{F}(k)$ & \textbf{Statistic} & \textbf{AUC} \\
\midrule
$\Phi_0$ raw                & 1.0000 & Trailing newline & \textbf{0.024} \\
$\Phi_1$ no comments        & 1.0000 & Lines per file   & 0.228 \\
$\Phi_2$ $+$ whitespace     & 1.0000 & Docstring        & 0.283 \\
$\Phi_3$ $+$ AST round trip & 1.0000 & Shebang          & 0.432 \\
$\Phi_4$ $+$ identifiers    & 1.0000 & Non-ASCII        & 0.432 \\
\midrule
\textbf{Features deleted} & $\mathcal{F}^{-}(m)$ & Mean line length & 0.538 \\
10 / 50                   & 1.0000 & \texttt{\_\_main\_\_} guard & 0.860 \\
200 / 1000                & 0.9971 / 0.9913 & & \\
\bottomrule
\end{tabular}
\end{table}

\subsection{Evasion and Graph Coverage}
\label{sec:scope}

We bound the verifier's scope by measurement rather than assertion, applying
seven source transformations to the 180 positives and reporting retention over
the applicable, still-parsing subset. Alias renaming, conditional import,
function-scoped import and a misleading vendor comment each retain 100\%
detection, whereas dynamic \texttt{\_\_import\_\_}, \texttt{importlib} and a
wrapper module each retain 0\%. The first group is within scope, and the second
defines what the method does not cover. The comment decoy fails because the
target is path-derived, which is both a strength against text manipulation and
the source of the metadata dependence noted earlier.

Recall is bounded by the binding table, which covers 7 of 24 mainstream Python OT
packages and leaves \lib{pylogix}, \lib{pyModbusTCP}, \lib{modbus\_tk},
\lib{minimalmodbus}, \lib{opcua}, \lib{asyncua}, \lib{OpenOPC}, \lib{pyads} and
\lib{pydnp3} among others uncovered, so a script using one of them is silently an unaudited negative (a potential false negative) rather than a
verified pass. All five vendor-lock edges carry a documented source.
Modbus-TCP is treated as vendor-neutral and never yields a witness, and a
protocol can be vendor-neutral while a device-family, function-code or
register-map assumption remains invalid, so vendor identity is a necessary but
not sufficient compatibility check.

Finally, the mismatch rate by supplied-context level is 14.4\% (low,
$n{=}340$), 14.2\% (medium, $n{=}316$) and 37.9\% (high, $n{=}227$) under the
reading in which an ``easy'' task supplies more target information, and reverses
under the opposite reading. We have not verified which reading the generating
study intends and assert no direction.

\section{Discussion}
\label{sec:discussion}
 
Two of the three claims survive scrutiny, since under the encoded bindings and
the corpus metadata roughly one machine-generated script in seven uses a library
not documented as supporting its declared target, with that library on the
communication path, and most such scripts lie outside an existence baseline, all
computed without labels or a trained model.
 
The third does not survive, because the provenance classifier's accuracy, and
the improvement attributed to grounding, cannot be read as authorship detection.
The construction is standard, with generated code from one collection and human
code from public repositories labelled by origin, yet the collections differ in
ways unrelated to authorship, since the human class is 75\% library and test
code set against entirely standalone scripts and the serialisation differed, so
a single byte suffices. Both confounds are invisible to usual reporting, and
work reporting macro-\Fone{} 0.98 under stratified cross-validation with
leave-one-model-out as a robustness check would pass review while measuring
nothing about authorship, as our own leave-one-model-out figure of 1.000 for
character $n$-grams shows. This is the code analogue of hypothesis-only baselines
in natural-language inference~\cite{gururangan2018}, within the pattern
documented for security machine learning~\cite{arp2022,pendlebury2019} and
leakage generally~\cite{kapoor2023}. The episode motivates three cheap sanity
checks for such studies, a single-statistic AUC over a handful of surface
features, a character $n$-gram baseline where any figure at or above the
semantic model signals a failed check, and a normalisation ladder that shows
which channel carries the signal.
 
The audit rests on metadata and partial coverage, which bound its claims. Target
vendors come from file naming, defensible metadata rather than verified intent, so
the rate depends on that channel and 44 flagged scripts name multiple vendors. We
release a blind
two-rater worksheet over all positives and 40 sampled negatives, and the
adjudicated rate remains future work, while only 28 of 190 human files are
adjudicable, so a matched corpus of human-written standalone OT scripts is the
binding constraint on any prevalence comparison. The verifier is as complete as
its bindings, exercises one context, checks vendor identity as a necessary but
not sufficient condition, and addresses an error made by a non-adversarial
generator rather than an adversary who has read this paper. It establishes
neither safety nor maliciousness, so deployment statements are feasibility
claims, and the earlier operating-cost argument for interpretability remains a
hypothesis until review cost is measured with analysts. On dual use, the bindings
are public and our contribution is their organisation, though we decline the
earlier assertion that an adversary gains nothing, since a mismatch checker can
also remove mismatches before deployment. We judge publication net-positive and
release the graph, evidence table, worksheet, splits, environment and pipeline,
but no executable attack scripts or live-controller tooling.
\section{Conclusion}
\label{sec:conclusion}

\textsc{Proteus-OT} decides metadata-conditioned, import-level vendor--protocol
incompatibility from source text and a documented binding table, without labels
or a trained model. The mismatch appears in 14.2\% of machine-generated scripts
in this corpus when the target is unambiguous and the offending library reaches a
communication call, rising to 20.4\% under the inclusive definition, and in none
of the 28 adjudicable human controls, with each positive carrying an inspectable
witness. On the same corpus, provenance classification separates machine from
human code almost perfectly but by construction rather than by authorship, with a
trailing byte alone sufficing, so accuracies reported for generated-code
detection on repository-labelled corpora reflect corpus construction and call for
the artifact audit demonstrated here. Where a domain supplies hard documented
constraints, grounding a check in them yields a decision reproducible by
construction. Future work should adjudicate the positives, broaden the graph
beyond its 7-of-24 coverage, exercise conditional context, and build the matched,
identically-serialised control corpus the provenance question requires.

\end{document}